\documentclass[prc,a4paper,preprint,showpacs,byrevtex,onecolumn]
{revtex4}
\everymath={\displaystyle}
\usepackage{rotating}
\def\1{\mbox{l\hspace{-0.53em}1}}

\def\r{{\bf r}}

\def\s{{\bf s}}

\def\sig{{\bf \sigma}}
\def\sig_1{{\bf \sigma_1}}

\def\beq{\begin{equation}}
\def\eeq{\end{equation}}
\def\bea{\begin{eqnarray}}
\def\eea{\end{eqnarray}}

\begin{document}
\title{Evolution of nuclear shells with the Skyrme density dependent
interaction}

\author{D. M. \surname{Brink $^{a}$}}
\email[E-mail: ]{thph0032@herald.ox.ac.uk}
\author{Fl. \surname{Stancu} $^{b}$}
\email[E-mail: ]{fstancu@ulg.ac.be}
\affiliation{$^{a}$ University of Oxford, Rudolf Peierls Centre of Theoretical Physics , 
1 Keble Road, Oxford OX1 3NP, U.K. \\
$^{b}$ University of Li\`ege, Institute of Physics, B.5, Sart Tilman,
B-4000 Li\`ege 1, Belgium.}



\begin{abstract}
We present the evolution of the shell structure of nuclei in 
Hartree-Fock calculations using Skyrme's density-dependent effective  
nucleon-nucleon interaction. The role of the tensor part of the 
Skyrme interaction  
to the Hartree-Fock spin-orbit splitting in spherical
spin unsaturated nuclei is reanalyzed. 
The contribution of a finite range tensor force to the spin-orbit splitting
in closed shell nuclei is calculated.
It is found that the exact matrix elements of a Gaussian and of a 
one-pion exchange tensor potential
could be written as a product Skyrme's short range expression times
a suppression factor  which is almost constant for closed shell nuclei with
mass number $A\geq 48$. The suppression factor is $\sim 0.15$ for the
one-pion exchange potential.
\end{abstract}



\maketitle

\section{Introduction}
The shell structure is a distinctive feature of nuclei and is characterized 
by the existence of magic numbers which are a consequence of the spin-orbit interaction
\cite{MAYER,JENSEN}.  The  spin-orbit interaction
can be understood in a mean field approach which leads to a one-body
potential containing a central part and a spin-orbit part. In spin-saturated
nuclei the spin-orbit part stems from the spin-orbit 
nucleon-nucleon interaction.
In spin unsaturated nuclei there are additional contributions coming 
either from the exchange part of the central two-body force or from
the tensor force \cite{SBF,VB,BEINER}. In view of the recent progress related 
to the discovery
of exotic nuclei (neutron or proton rich) a major problem is to understand
how the shell structure evolves  from stable to unstable nuclei.
Presently there is much concern about the role of the tensor force
in the shell evolution and the structure of exotic nuclei 
\cite{OTSUKA1,OTSUKA2,BROWN,OTSUKA3,COLO,PIEKAREWICZ}.

In a previous work \cite{SBF} we  estimated the contribution of the 
tensor part of the Skyrme interaction to the Hartree-Fock spin-orbit
splitting in several magic nuclei and adjusted the strength of the tensor
force such as to obtain a good global fit. In the present paper we extend the 
previous study to exotic nuclei, most of which were unknown
at that time. This extension sheds a new light on the previous results.

The tensor term in the Skyrme interaction is written as a $\delta $-function 
in the internucleon separation multiplied by momentum dependent terms (Sec. 4). 
The momentum dependence takes the finite range of the interaction into account. 
Contrary to the view that it plays a minor role because of its $\delta$-type 
structure  \cite{OTSUKA1}, this interaction 
has the same effect as a finite size interaction due to its 
momentum dependence. We will show that the Skyrme interaction provides a good 
mechanism for describing the evolution of the shell structure in exotic nuclei.

Otsuka et al. \cite{OTSUKA1} have pointed out that the nuclear tensor force has a 
rather long range and that  the use of the energy density (\ref{Hr}) may not be 
justified. The goal of the present paper is to show that expressions (\ref{Hr})
together with (\ref{Wn}) and  (\ref{Wp}), 
given in Section 2 of this paper, can still be used to study the contribution of finite range tensor forces, even if the range is that of the one-pion exchange force. 
Shell gaps are mainly determined by the spin-orbit splitting of the states with 
highest $l$ in any shell and our study is restricted to these states. The spin-orbit 
splitting is less important in states with lower $l$ because it is hidden by 
pairing effects and other forms of configuration mixing.

In Section 2 we derive an expression for the leading contribution 
of the tensor force to the Skyrme energy density functional using some results 
from the paper of Negele and Vautherin \cite{Negele} on the density matrix 
expansion method. Section 3 presents some numerical calculations which show that 
the main effect of a longer range interaction is to introduce a suppression factor 
which is almost constant for all nuclei with mass number greater 
than $A \sim 28$. Section 4 recalls the expression of the tensor for a
short range tensor interaction.
In Sec. 5 we present results for single particle levels of Sn isotopes,
N = 82 isotones and Ca isotopes, where the tensor force considerably 
improves the agreement with the experiment when its parameters
are properly chosen.

The conclusion is that the Skyrme energy functional with the tensor force
is adequate to describe the evolution of shell effects.

\section{Contribution of  a short range tensor force}

The Skyrme parametrization of a short range tensor force leads to a contribution to the 
energy density
\beq
\Delta H(\r) = \frac{1}{2} \alpha[J^2_n(\r) + J^2_p(\r)] + \beta J_n(\r) J_p(\r)
\label{Hr}
\eeq
where the $J_q(\r)$ ($q = n,p$)  are spin-orbit densities
and $\alpha$ and $\beta$ are parameters defined in Section 5. 
They represent the combined effect of the tensor plus central exchange
interactions. If the radial wave 
functions depend only on the orbital angular momentum $l$ and not on $j$ then 
the spin-orbit densities are zero 
if both components of a spin-orbit doublet are filled. Then the energy density 
(\ref{Hr}) brings contributions only to spin unsaturated nuclei. 
Thus $\Delta H(\r)$ would 
be almost zero in $^{40}$Ca which is a double closed shell nucleus. It would be large for 
$^{208}$Pb which has spin unsaturated shells for both neutrons and protons. The energy
functional (\ref{Hr}) leads to a simple modification of the single particle 
spin-orbit potential for both protons and neutrons (see Section 5).  

The purpose of the present section is to derive the form of $\Delta H(\r)$ for a short 
range tensor interaction. We focus on the contribution of the neutron-proton interaction.
The starting point is a two-body tensor potential 
\beq
V_T(r) =  v_T(r)~ \vec{\tau_1} \cdot \vec{\bf \tau_2} 
~[~ \frac{1}{3 r^2}(\vec{\sigma_1} \cdot \vec{r})(\vec{\sigma_2} \cdot \vec{r})- 
\vec{\sigma_1} \cdot \vec{\sigma_2}]
\label{t1}
\eeq
like the one arising from one-pion exchange. The effect of the isospin dependence in 
Eq. (\ref{t1}) is to make $\beta \approx 2\alpha$. An interaction with no isospin 
dependence would have $\beta = 0$.

According to Negele and Vautherin \cite{Negele} the expectation value of the tensor 
interaction with Hartree-Fock wave functions is
\beq
\langle V_{T}^{np}\rangle =- \int d^3\r_1 d^3\r_2 v_T(\r_1-\r_2)
| {\vec \rho}_n(\r_1,\r_2).{\vec \rho}_p(\r_1,\r_2) |
\label{vt0}
\eeq
The expressions for the $nn$ and $pp$ contributions are similar but are each multiplied 
by a factor $1/2$ from the isospin dependence of (\ref{t1}). Negele and Vautherin give a 
factorization of the spin-density matrices for spherical nuclei in which sub-shells are 
either completely full or completely empty. It is
\beq
{\vec \rho}(\r_1,\r_2) = i(\r_1\times \r_2) \rho_1(\r_1,\r_2)
\label{eq6}
\eeq
where
\beq
\rho_{1}(\r_1,\r_2) = \pm \sum_{njl}
\frac{1}{ 2 \pi r_1^2 r_2^2} R_{njl}(r_1) R_{njl}(r_2) P'_{l}(\cos \theta ) .
\label{rho1}
\eeq
$P'_l$ is the derivative of the Legendre polynomial $P_l$, $\theta$ is the angle 
between the directions of $\r_1$ and $\r_2$ and the $\pm$ sign in 
Eq. (\ref{rho1}) stands 
for $j = l\pm \frac{1}{2}$. For a short range interaction $\theta \approx 0$ in 
Eq. (\ref{rho1}) and $P'_{l}(\cos \theta ) \approx l(l+1)/2$. This is the origin 
of the spin-orbit splitting factor in Eq. (6) of Ref. \cite{SBF} 
\beq
(2j+1)[j(j+1) -l(l+1) - 3/4]= \pm 2l(l+1)\quad {\rm if}
 \quad\ j=l\pm 1/2.
\label{sign}
\eeq
If the radial wave functions $R_{njl}(r)$ are the same for $j = l \pm 1/2$ then 
the contribution of a particular $l$-level to $\rho_1(\r_1,\r_2)$ vanishes if 
both $j$-components are either completely occupied or completely empty.

When the interaction $V_T(r)$ has a sufficiently short range the 
$\langle V^{np}_{T}\rangle$
simplifies to
\beq
\langle V^{np}_{T}\rangle = \frac{\pi}{6} \int  d^3 \r J_n(r)J_p(r) \int v(s) s^4 d s .
\label{short}
\eeq
where 
\beq
J_{q}(r) = 2 r \rho_{1q}(\r,\r)
\eeq
which is  the spin-orbit density in Eq. (6) of Ref. \cite{SBF}. Equation (\ref{sign}) shows 
that $J_{q}(r) >0$ when the lower component of a spin-orbit doublet is being filled and 
goes to zero when both components are filled.

\section{A finite range suppression factor}

The analysis in the present section  shows that, for a tensor interaction with a 
range of the order of the one pion exchange potential  and for single particle 
states with the largest $l$ for a given $A$,  the effect of the finite range 
interaction is to multiply Eq. (\ref{short}) by a simple suppression factor which 
is almost the same for any nuclei with mass number greater than $A= 28$. 
As a consequence one should be able to parametrize the contribution of a tensor force 
to the energy density by the simple form in Eq. (\ref{Hr}) with values of $\alpha$ 
and $ \beta $ which are constant for all nuclei. This means that the original 
short range Skyrme density dependent form used in \cite{SBF} remains entirely 
valid and that
finite range effects can be incorporated by using suitable values of 
$\alpha$ and $\beta$ (see Section 5).

We start with Eq. (\ref{vt0}) and make  a change of variables in the expression 
for $\langle V^{np}_{T}\rangle$ which becomes
\beq
\langle V^{np}_{T} \rangle = 8 \pi^2 \int_0^{\infty}F(r) dr ~,
\eeq
where 
\beq
F(r) = \int_0^{2r} ds_r \int_0^{\pi} d\theta~ \sin^3  \theta \
 v_T(\s)(r_1r_2)^4  \rho_{1n}(\r_1,\r_2)\rho_{1p}(\r_1,\r_2),
\label{vt1}
\eeq
with $r_1 = r + s_r/2$, $r_2 = r - s_r/2$ and $\theta $ the angle between $\r_1$ 
and $\r_2$. There are three other angles which have been integrated out to give a 
factor $8\pi^2$. The formula contains the
factor $|\r_1 \times \r_2|^2 = (r^2 - s_r^2/4)^2 \sin^2 \theta $.
The spin densities $\rho_{1q}$ are defined in Eq. (\ref{rho1}).
The squared distance $|\s|$ between the points $\r_1$ and $\r_2$ is 
\beq
|\s|^2 = |\r_1 -\r_2|^2 =  s_r^2 + 4(r^2 - s_r^2/4)~\sin^2 \frac{\theta}{2}~,
\label{sr}
\eeq

For states with maximum $l$ in any shell the function $F(r)$ has a single peak at $r_m$
near the maximum of the radial wave functions. The short range approximation
to $F(r)$, denoted by $F_0(r)$,
holds when the range of the interaction is much less than
$r_m$. The important values of $s_r$ are much less than $r_m$ and the angle integral has contributions only 
from small values of $\theta $. Then $F(r)$ is replaced by
\beq
F_0(r) = \int_0^{\infty} ds_r\int_0^{\infty} d\theta \ \theta^3 \
 v_T(\s)r^8  \rho_{1n}(\r,\r)\rho_{1p}(\r,r),
\label{short2}
\eeq
with $|\s|^2 = {s_r}^2 + r^2\theta^2$. 
The short range approximation  (\ref{short}) to $\langle V^{np}_{T} \rangle $ can be obtained 
from (\ref{short2})
by using the relation (\ref{rho1}). 

The numbers in Table 1 are calculated with oscillator radial wave function
\[
R_l(r) =N_l r^{l+1}\exp \left(-\frac{r^2}{2b^2}\right).
\]
They have a maximum at $r_m = b\sqrt{l+1}$. 
The radial suppression factor $S(r)$ and the total suppression factor $S$ are defined by
\beq
S(r) = \frac{F(r)}{F_0(r)}, \qquad S = \frac{\int F(r) dr}{\int  F_0(r) dr}~.
\label{SmS}
\eeq

There is a  simple approximation $I(l,r/a)$ for the suppression factor $S^G(r)$
for a Gaussian interaction $V(r) = V_0 \exp(-r^2/a^2)$ which is given by
\beq
I(l,r/a )   = \left(\frac{4r^2/a^2}{l(l+1) + 4 r^2/a^2}\right)^2
\left[ \frac{1}{1+ a^2/b^2}\right]^{1/2}
\label{approxI}
\eeq
with  $l=(l_n +l_p)/2$.
The form of Eq. (\ref{approxI}) arises from the replacement of $r^2- {s^2_r}/4$ 
and of $r_1r_2$ by $r^2$  in Eqs. (\ref{vt1})
and (\ref{sr}), an approximation which is valid when $s_r \sim a <<r$. Then the 
integral (\ref{vt1}) separates into a product of an angle integral and an 
integral over $s_r$ for a Gaussian interaction. The first factor is independent of the 
radial wave functions.

 Values of $I(l,r_m/a )$ for several closed shell nuclei
are given in Table 1. The range $a=1.2$ fm is taken from \cite{OTSUKA3}.
 There is quite a strong dependence on $r$ for fixed $l$ but 
$I(l,r/a)$ is almost constant at $r=r_m$ because  
$l(l+1)a^2 /r_m^2 $ does not change much for all the nuclei considered.
The table also gives values of $S^G(r_m)$ and the total suppression factor $S^G$ calculated by numerical integration. 
The approximate formula (\ref{approxI}) for $S^G(r_m)$ is accurate to within 5\%. 
The results indicate that, for a Gaussian potential  and for states with 
maximum $l$, one can use the short range approximation (\ref{short})
with a reduced interaction  strength.

\[
\begin{array}{cccccccc}
 A & l  & r_m(fm) &  I(L,r_m/a)& S^G(r_m)&   S^G   &  S^Y(r_m )&  S^Y  \\
28 & 2  & 3.08 &     0.550     &   0.579 &   0.459 &  0.197 &   0.159 \\
48 & 3  & 3.89 &     0.515     &   0.521 &   0.436 &  0.171 &   0.147 \\
90 & 4  & 4.83 &     0.511     &   0.507 &   0.440 &  0.166 &   0.146 \\
132 & 4,5 & 5.43 &   0.520    &    0.511 &   0.449 &  0.166 &   0.147 \\
208 & 5,6& 6.34 &    0.516     &   0.504 &   0.452 &  0.164 &   0.145 
\end{array}
\]
\begin{center} Table 1: Values of the suppression factors $I(l,r_m/a)$, $S^G(r_m)$, 
$S^G$, \\ $S^Y(r_m)$ and $S^Y$ for the
Gaussian and one-pion exchange potentials \\ for various values of $A$ and $l$. \end{center}

The final two columns of the table give values of $S^Y(r_m)$ and $S^Y$ for 
the tensor part of a Yukawa (one-pion exchange) potential which has a longer range 
and a form factor with radial dependence  
$v_T(x) = V_0\exp(-x)(1/x +3/x^2 +3/x^3)$, where $x=\mu s$ 
with $\mu$ = 0.70  fm$^{-1}$. It has a $1/s^3$ 
singularity at $\s=0$ but this is cancelled by the $\sin^3 \theta $ factor in the integral
for the matrix element. 
The  suppression factors $S^Y(r_m)$ and  $S^Y$, calculated by numerical integration, are
almost constant. This shows that it is reasonable to 
use the Skyrme parametrization (\ref{Hr}) to study the contribution of 
the tensor force to spin-orbit splittings for states with maximum $l$
even for the one-pion exchange potential. 

 Our calculations show that the suppression factor $S(r)$
is an increasing function of $r$ which goes to zero as $r\rightarrow 0$ 
and to 1 for larger $r$. 
The numbers in the table show that the total suppression factor $S$ is less than
the suppression factor evaluated at $r_m$ for both the Gaussian and one-pion
exchange potentials. 

Early calculations showed that the Yukawa one-pion exchange potential
play an important role in describing the $^{208}$Pb levels 
\cite{SAVUSHKIN}.
Note that the results given above for the Gaussian potential shape are important
in view of the fact that such interactions are used in shell model calculations
(cf. Otsuka et al. \cite{OTSUKA3}).
The main difference between the values for $S^G(r_m)$ and $S^Y(r_m)$ in Table 1
is due to the range of the Gaussian interaction. The values become very similar
for all the nuclei in the table if the range of the Gaussian interaction is 
increased from 1.2 fm to 2.1 fm.

\section{The tensor part of the Skyrme interaction}

The parameters of the Skyrme interaction were originally determined 
in Hartree-Fock calculations to reproduce the 
total binding energies and charge radii of closed-shell nuclei \cite{VB}. 
Further extensive calculations were made 
later \cite{BEINER}. Several improved parameter sets were found.    
They differ mainly  through the single particle spectra. In the present paper 
as in our previous work, we shall use the parameter set SIII which
gives good overall single particle spectra. In Ref. \cite{SBF} a tensor 
force was added and a range of its strength was found such as to maintain a
good quality of the single particle spectra of $^{48}$Ca,$^{56}$Ni, $^{90}$Zr
and $^{208}$Pb.   

As in Ref. \cite{SBF}, in the configuration space the tensor interaction has the following form
\begin{eqnarray}
V_{T} = \frac{1}{2} T \{ [(\vec{\sigma_1} \cdot \vec{k'})  
(\vec{\sigma_2} \cdot \vec{k'}) - \frac{1}{3} k'^2 (\vec{\sigma_1} \cdot
\vec{\sigma_2})] \delta{(\vec{r_1}-\vec{r_2})}  \nonumber \\
+ \delta{(\vec{r_1}-\vec{r_2})} [(\vec{\sigma_1} \cdot \vec{k}) 
(\vec{\sigma_2} \cdot \vec{k})      
 - \frac{1}{3} k^2 (\vec{\sigma_1} \cdot
\vec{\sigma_2})] \delta{(\vec{r_1}-\vec{r_1})} \} \nonumber \\
+U \{(\vec{\sigma_1} \cdot \vec{k'}) \delta{(\vec{r_1}-\vec{r_2})}
 (\vec{\sigma_1} \cdot \vec{k})
 -\frac{1}{3} (\vec{\sigma_1} \cdot \vec{\sigma_2})
 [\vec{k'} \cdot \delta{(\vec{r_1}-\vec{r_2})} \vec{k}] \}.  
\end{eqnarray}
The parameters $T$ and $U$ measure the strength of the tensor force in 
even and odd states of relative motion.

\section{Results}

The analysis in Sections 2 and 3 show that the simple form (\ref{Hr}) 
is a good approximation to the contribution of the tensor forces to 
the energy density. Values of $\alpha$ and $\beta$ can be taken to be 
constant for states with maximum $l$ in nuclei with $A\geq 48$ even for 
forces with a range of the one pion exchange potential.

Both the central exchange and the tensor interactions
give contributions to the spin-orbit single particle potential to be added
to the spin-orbit interaction. The additional contribution are \cite{SBF}

\begin{equation}
\Delta W_n = (\alpha J_n + \beta J_p) \vec{\ell} \cdot \vec{s}
\label{Wn}
\end{equation}

\begin{equation}
\Delta W_p = (\alpha J_p + \beta J_n) \vec{\ell} \cdot \vec{s}
\label{Wp}
\end{equation}
with $\alpha = \alpha_T + \alpha_c$ and  $\beta = \beta_T + \beta_c$.
For the Skyrme SIII interaction the parameters of the central exchange part 
are \ \cite{BEINER}
\begin{equation}
\alpha_c = \frac{1}{8} (t_1 - t_2) = 61.25~ \mathrm{MeV~ fm^5},
~~~~~~\beta_c = 0~,
\end{equation}
where $t_1$ and $t_2$ are two of the Skyrme interaction parameters.
In terms of the tensor parameters $T$ and $U$ one has 
\begin{equation}
\alpha_T = \frac{5}{12} U, 
~~~~~~\beta_T = \frac{5}{24} (T + U).
\end{equation}
Equations (\ref{Wn}) and (\ref{Wp}) imply that the mechanism invoked by 
Otsuka et al. is intrinsic to the Skyrme energy density formalism. These 
equations show that the filling of proton (neutron) levels influences the spin-orbit
splitting of neutron (proton) levels whenever $\beta \neq 0$. 
The normal spin-orbit single particle potential is
\begin{equation}
V_{so} =  W_0 \frac{1}{r} (\frac{d \rho}{d r} + 
\frac{d \rho_q }{d r}) \vec{\ell} \cdot \vec{s}\qquad {\rm with }\qquad \frac{d \rho}{dr}<0.
\end{equation}
When $\beta$ is positive the neutron (proton) spin-orbit splitting is reduced as 
protons (neutrons) fill a $j = l+1/2$ level because $J_{p(n)} >0$. This effect 
is clearly seen in Fig. 4 of Otsuka et al. \cite{OTSUKA2}.

In Ref. \ \cite{SBF} we searched for sets of parameters 
$\alpha$ and $\beta$ which simultaneously fit absolute values of single particle 
levels in the closed shell nuclei  $^{48}$Ca, $^{56}$Ni, $^{48}$Zr and 
$^{208}$Pb. We found that the common optimal values were located in a 
right angled triangle 
with sides $\alpha$ = - 80 MeV fm$^5$, $\beta$ = 80 MeV fm$^5$
and hypotenuse $\alpha + \beta = 0$. 
Here we relax these constraints and try to analyze single particle energies
some  nuclei far from the stability line.  The experimental data   did not exist in 1977
when we discussed the global fit for closed shell nuclei \cite{SBF}. 
Our present choice of parameters is guided by the recent results
of Ref. \cite{COLO} on the Z = 50 isotopes and N = 82 isotones which
were analyzed in a HF + BCS approach based on the Skyrme interaction 
SLy5 \ \cite{CHABANAT} with refitted values of $T$ and $U$
plus a pairing force. 
To see whether or not one can obtain the correct trend in the evolution of 
single particle levels we look at energy differences between them. These differences 
can give a clear indication of the formation of closed shells from the size of the gaps.
Absolute values of single particle energies depend not only on the tensor, 
but also on other parts of the Skyrme interaction. Here we are not concerned with
making the best fits to absolute energies.

In the present paper we still use the SIII version of the Skyrme 
interaction \ \cite{BEINER} for comparison with the previous work.  
We maintain the conditions 
$\alpha < 0$ and $\beta > 0$ but take values outside the triangle
found before which are not inconsistent with the previous findings 
\ \cite{SBF}.  We show that  values $\alpha_T$ = - 180 MeV fm$^5$
and $\beta_T$ = 120 MeV fm$^5$,  or equivalently 
$\alpha$ = - 118.75 MeV fm$^5$ and $\beta$ = 120 MeV fm$^5$, 
give a reasonably good fit to Z = 50 isotopes and N = 82 isotones.
These values are similar to the ones fitted by Brown et al. \cite{BROWN} in a recent paper. 
For a more general orientation we also discuss the role of this
parametrization on Ca isotopes.

We conclude this section with some remarks on $^{208}$Pb and $^{90}$Zr. 
The proton h$_{11/2}$ and neutron i$_{13/2}$ in $^{208}$Pb are 
filled and $J_p$ and $J_n$ are both positive with comparable magnitudes. 
Because $\alpha \approx -\beta$ we have 
$\Delta W_n \approx \Delta W_p \approx 0$ and the tensor forces 
hardly change the spin-orbit splitting. The situation is different for 
$^{90}$Zr. There $J_p = 0$ and the effect of the tensor forces is to 
increase the spin-orbit splitting for neutrons and reduce it for protons. 
The shell gaps for protons and neutrons are both increased significantly 
and the stability of the double closed shell at $^{90}$Zr is enhanced.

\subsection{Sn isotopes} 

\begin{figure}
\begin{center}
\includegraphics[width=12cm]{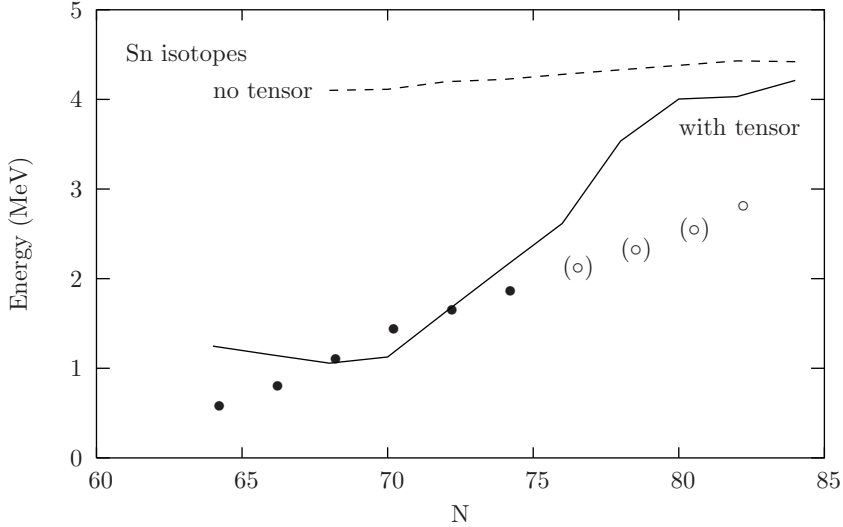} \\
\caption{The proton single particle energy 
difference between $1h_{11/2}$ and $1g_{7/2}$ in Sn isotopes (Z = 50, N = 64-82)
calculated
without and with tensor force $\alpha$ = - 118.75 MeV fm$^5$,
$\beta$ = 120 MeV fm$^5$. 
Data points are from Ref. \cite{SCHIFFER}. Solid dots give information from
transfer reactions.  Open circles are obtained from methods less sensitive
to the single particle nature. The parentheses indicate less certain or 
indirect assignments.}
\label{Sn}
\end{center}
\end{figure}

Fig. \ref{Sn} shows the HF results for the proton
single particle energy difference 
between $1h_{11/2}$ and $1g_{7/2}$ in Sn isotopes (Z = 50, N = 64-82) 
with and without tensor force.
One can see that the effect of the tensor force is indeed important.
The experimental pattern is satisfactorily reproduced with this
simple approach. In the more sophisticated HF+BCS calculations of  
Ref. \cite{COLO} the theoretical results 
beyond  $^{126}$Sn  are better.
However in that region the experimental situation is 
less certain 
because the corresponding values have been assigned using methods
which are less sensitive to the single particle nature of 
the levels \cite{SCHIFFER}. For the double magic nucleus $^{132}$Sn 
the effect of the tensor force and of the central exchange part cancels out
because $J_p \approx J_n$.
For isotopes with Z = 56-62 the comparison with the experiment is 
not possible because the $1h_{11/2}$ level becomes unbound in these 
calculations.

\subsection{N = 82 isotones}

\begin{figure}
\begin{center}
\includegraphics[width=12cm]{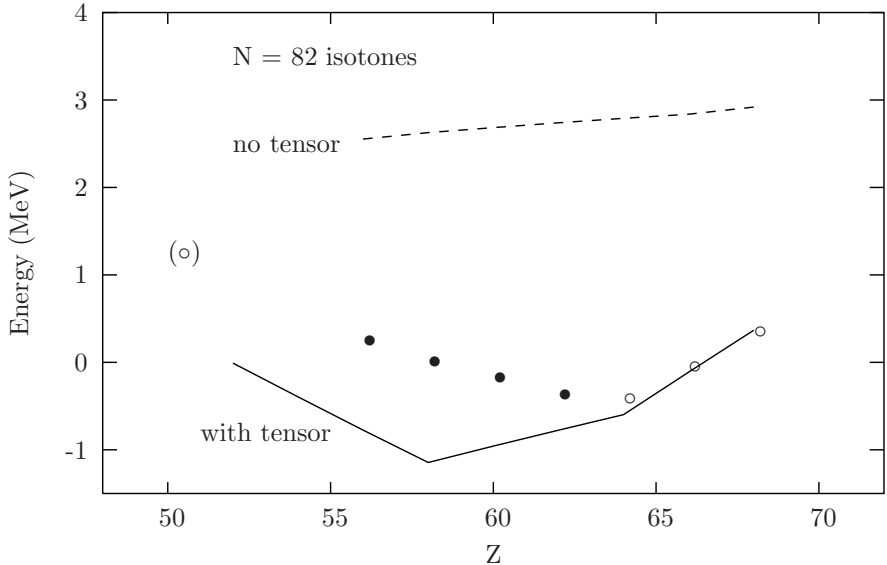} \\
\caption{The neutron single particle energy 
difference between $1i_{13/2}$ and $1h_{9/2}$ in N = 82 isotones calculated
with  and without  tensor force. 
Data points are from Ref. \cite{SCHIFFER}. 
Solid dots give information from
transfer reactions.  Open circles are obtained from methods less sensitive
to the single particle nature. The parentheses indicate less certain or 
indirect assignments.}
\label{ISOTONES}
\end{center}
\end{figure}

In Fig. \ref{ISOTONES} we present neutron single particle energy 
differences between $1i_{13/2}$ and $1h_{9/2}$ in N = 82 isotones calculated
with  and without  tensor force and compare them with data from  
Ref. \cite{SCHIFFER}. Again the role of the tensor force is considerable,
bringing down  the energy difference $e(1i_{13/2}) - e(1h_{9/2})$
close to the best known experimental values. For Z $\leq$ 50 and for Z $\geq$ 70 the $1h_{9/2}$
level becomes unbound both with and without tensor force.

\subsection{Ca isotopes}

\begin{figure}
\begin{center}
\includegraphics[width=12cm]{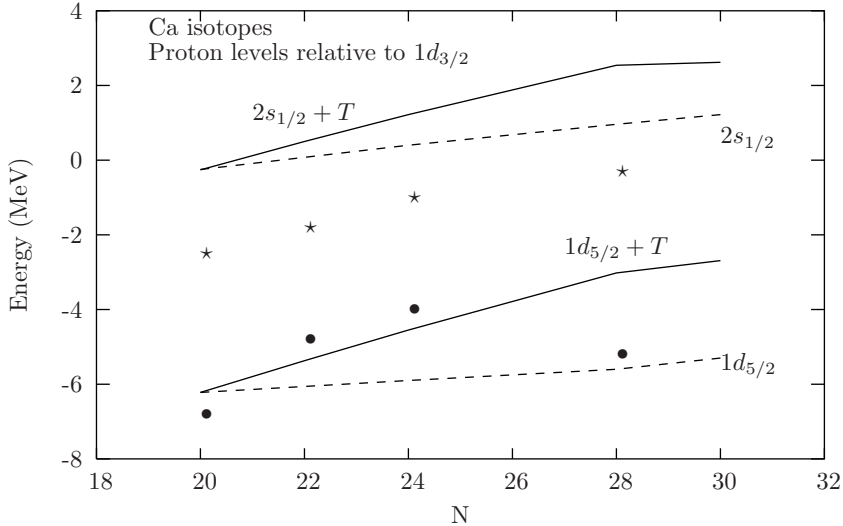} \\
\caption{The proton single particle energies in Ca isotopes 
relative to $1d3/2$ level, calculated
with tensor force ($nlj + T$): $\alpha$ = -118.75 MeV fm$^5$, $\beta$ = 120 
MeV fm$^5$  and without tensor force 
($nlj$): $\alpha$ = 61.25 MeV fm$^5$, $\beta$ = 0. 
Data points are from Ref. \cite{COTTLE}: solid dots 
for $1d5/2$ and stars for $2s1/2$.}
\label{Ca}
\end{center}
\end{figure}

As the Skyrme interaction  SIII was fitted to closed shell nuclei 
it has the peculiarity that the
central exchange interaction produces some undesirable effects in the middle
of a shell. The predicted single particle levels have the wrong order 
when compared with the experimental levels and wrong levels are occupied
\ \cite{BEINER}. This happens in the absence of the tensor interaction, 
but when a tensor interaction with adequate parameters is added 
the problem is solved. In particular
the parameters $\alpha$ = - 118.75 MeV fm$^5$ and $\beta = 120$ MeV fm$^5$ remove this
anomaly in $^{50}$Ca.
The reason is a considerable increase of the spin-orbit in the $1f$ shell
which shifts  the $1f_{7/2}$ above the $2p$ levels.
However the anomaly persists  for $\alpha$ = 0 and $\beta $ = 80 MeV fm$^5$
located at the edge of the above mentioned triangle.

In addition, from Fig. \ref{Ca} one can see that
the effect of the tensor interaction is important and improves   
the spin-orbit splitting in the $1d$ shell. 
In the $2s_{1/2}$ shell the trend is correct but the theoretical results
are above the experimental points, with or without tensor. The pattern is quite 
similar to that obtained in Ref. \cite{OTSUKA1} in a shell model approach.

\section{Conclusions}

The short range approximation for the contribution of a tensor force 
to the spin-orbit splitting in nuclei was 
studied in sections 2 and 3 for both a Gaussian and a Yukawa (one-pion exchange) 
interactions for states with maximum $l$ in any shell for several nuclei with
mass number between A = 28 and A = 208. It was shown that the 
exact matrix elements of the one-pion exchange tensor potential
could be expressed as a product of the short range expression (\ref{short}) and
a suppression factor $S^Y \approx 0.147$ which is almost constant for nuclei 
with mass number $A\geq 48$. It is only slightly larger, 
$\emph{i. e.}~ S^Y \approx 0.16$ for nuclei near $^{28}$Si.
Thus the short range formulae 
(\ref{Hr}), (\ref{Wn}) and (\ref{Wp}) with constant $\alpha$ and $\beta$  should 
give qualitatively good results   for a Yukawa one-pion exchange potential. 

We have made a new fit to the parameters $\alpha$ and $\beta $ in the 
parametrization (\ref{Wn}) and (\ref{Wp}) of the tensor contribution 
to the spin-orbit coupling using data on $Z=82$ isotopes and $N= 82$ 
isotones. The tensor force makes a dramatic difference to the single 
particle energy difference between the h$_{11/2}$ and g$_{7/2}$ single 
particle levels. A similar situation holds for the energy difference 
between the i$_{13/2}$ and h$_{9/2}$ single particle levels in $N=82$ 
isotones. In both cases the calculation with the addition of the tensor 
force give a good description of the experimental data. The  case with Ca isotopes
is similar to $^{90}$Zr. The tensor force reduces the spin-orbit splitting 
for p  rotons and increases it for neutrons. This brings the order of single 
particle into a better agreement with experiment.

The mechanism observed by Otsuka and collaborators \cite{OTSUKA1,OTSUKA2, OTSUKA3} 
that the filling of neutron levels
influences the proton spin-orbit splitting and vice-versa is intrinsic to 
 the Skyrme energy density approach and is very simple in that theory. 
In addition, with the Skyrme density formalism, one can easily study
the combined contribution of the central exchange and tensor NN interactions
to the spin-orbit potential.




\begin{thebibliography}{11}

\bibitem{MAYER} M. G. Mayer, Phys. Rev.  {\bf 75}, 1969 (1949).
 
\bibitem{JENSEN} O. Haxel, J. D. H. Jensen and H. E. Suess,
Phys. Rev.  {\bf 75}, 1766 (1949).
  
\bibitem{SBF} Fl. Stancu, D. M. Brink and H. Flocard, 
Phys. Lett. {\bf B68},  108 (1977).

\bibitem{VB} D. Vautherin and D. M. Brink, Phys. Lett. {\bf B32}, 149 (1970);
Phys. Rev. {\bf C5}, 626 (1972).

\bibitem{BEINER} M. Beiner, H. Flocard, Nguyen van Giai and P. Quentin,
Nucl. Phys.  {\bf A238}, 29 (1975).

\bibitem{OTSUKA1} T. Otsuka, T. Suzuki, R. Fujimoto, T. Matsuo, D. Abe,
H. Grawe and Y. Akaishi, Acta Phys. Polon. {\bf B36},  1213 (2005).

\bibitem{OTSUKA2} T. Otsuka, T. Suzuki, R. Fujimoto, 
H. Grawe and Y. Akaishi, Phys. Rev. Lett. {\bf 95}, 232502 (2006).

\bibitem{BROWN} B.A. Brown, T. Duguet, T. Otsuka, D. Abe and T. Suzuki, 
Phys. Rev. {\bf C 74},
061303 (2006).

\bibitem{OTSUKA3} T.Otsuka, T. Matsuo and D. Abe, Phys. Rev. Lett. {\bf 97}, 162501 (2006).

\bibitem{COLO} G. Col${\grave {\rm o}}$, H. Sagawa, S. Fracasso and 
P-F. Bortignon, Phys. Lett. {\bf B646}, 227 (2007).

\bibitem{PIEKAREWICZ} J. Piekarewicz, J. Phys. {\bf G 34}, 467 (2007).

\bibitem{Negele} J.W. Negele and D. Vautherin, Phys. Rev. C {\bf 5}, 1472 (1972).




\bibitem{SAVUSHKIN} L. N. Savushkin and V. N. Fomenko, 
Sov. J. Nucl. Phys. {\bf 28}, 29 (1978).

\bibitem{CHABANAT} E. Chabanat, P. Bonche, P. Haensel, J. Meyer and R.Schaeffer,
Nucl. Phys. {\bf A635}, 231 (1998).

\bibitem{SCHIFFER} J. P. Schiffer et al., Phys. Rev. Lett. {\bf 92}, 162501
(2004).

\bibitem{COTTLE} P. D. Cottle and K. W. Kemper, 
Phys. Rev. {\bf C58}, 3761 (1998).

\end{thebibliography}
\end{document}